\documentstyle[multicol,aps,prb,psfig]{revtex}
 
\def\beginwide{
        \end{multicols} \vspace*{-0.5cm} \noindent
        \rule{3.5in}{.1mm}\rule{.1mm}{5mm} \widetext \medskip }
\def\beginwidetop{
        \end{multicols} \vspace*{-0.5cm} \noindent
        \widetext \medskip }
\def\endwide{
        \hspace*{3.5in}~\rule[-5mm]{.1mm}{5mm}
\hspace*{-1mm}\rule{3.53in}{.1mm}
        \begin{multicols}{2} \vspace*{-1.0cm} \noindent }
\def\endwidebottom{
        \begin{multicols}{2} \vspace*{-1.0cm} \noindent }

\draft
\begin{document}

\title{Role of the superconducting gap opening on vertex corrections} 
\author{E.  Cappelluti$^{1}$, 
C. Grimaldi$^{2}$ and L. Pietronero$^{2,3}$}

\address{$^{1}$Max-Planck-Institut f\"{u}r Festk\"{o}rperforschung,
Heisenbergstrasse  1,
D-70569 Stuttgart, Germany}
\address{$^{2}$Dipartimento di Fisica, Universit\'{a} di Roma ``La
Sapienza", 
Piazzale A.  Moro, 2, 00185 Roma, Italy \\
and Istituto Nazionale Fisica della Materia, Unit\'a di Roma 1, Italy}
\address{$^{3}$ ICTP, P.O.  Box 586, 34100
Trieste, Italy}
\maketitle 
\medskip
\begin{abstract}

The nonadiabatic effects due to the breakdown of Migdal's theorem
in high-$T_c$ superconductors are strongly affected by the
opening of the superconducting gap. Here we report how the
momentum-frequency dependence of vertex corrections is 
modified by the gap. A general effect is that the positive region
of the vertex corrections is increased leading to an enhancement
of their relevance. This has a number of physical consequences
that we discuss in details in relation to specific experiments.
\end{abstract}
\vskip 2pc 
\begin{multicols}{2}

\section{Introduction}
A common peculiar characteristic of many unconventional
high-$T_{c}$ superconductors (cuprates, 
$A_{3}C_{60}$ compounds,...) is the narrowness of the electronic
bands crossing the Fermi level, leading to Fermi energies $\epsilon_F$
of the same order of magnitude of the phonon energies $\omega_{ph}$
($\omega_{ph}/\epsilon_F \div$ 0.2 - 0.3).
\cite{plakida}
In such a situation, the adiabatic assumption 
($\omega_{ph}/\epsilon_F \ll$ 1) on which
Migdal's theorem\cite{migdal} rests cannot be used anymore
to justify the omission of
vertex corrections in a diagrammatic theory.
Motivated by this observation, a renewed interest has recently
arose about the possible effects due to inclusion of the vertex corrections in the electron-phonon (or any kind
of bosonic mediator) interaction.
This task is even more relevant for the purely electronic
interaction models
(Hubbard, of $t-J$), where no justification exists at all
to neglect such corrections.\cite{schrieffer}

In this perspective, in previous papers, we have performed an 
intensive study of the
vertex corrections in the normal state, and, afterwards,
we have generalized 
the conventional Migdal-Eliashberg theory in order to
include the first nonadiabatic corrections due to the
breakdown of Migdal's theorem.\cite{PSG,GPSprl,CP} 
We have investigated the effect of the
Migdal corrections on different quantities related to both the
superconducting and the normal state properties, as for instance the 
superconducting critical temperature, or the isotope effect on the
effective electron mass.\cite{GCP} 
It was shown that an important feature
of these corrections is the non trivial
structure in momenta and frequencies space, so that the actual
relevance of their inclusion can, or cannot, affect in a
significant way the properties of the system depending on
the effective parameters of the systems.
In particular, the critical temperature $T_c$ is found to be sensitive
to the momentum transfer of the electron-phonon interaction,
and it would be interesting to evaluate the effect of the
vertex corrections to the gap-$T_c$ ratio.
Moreover, a negative isotope effect is predicted on
the effective electron mass $m^*$ in the normal state. 
This is in quite good agreement with recent experiments that 
show a large isotope effect on $m^*$.\cite{zhao,muller}
However, this result is based on measurements performed in the
superconducting state, where a nonadiabatic generalization
of Migdal-Eliashberg theory to include vertex corrections is
still
missing, and no prediction on the effect of vertex
corrections can be yet done.\cite{freericks}

As a first step to overcome this lack, we are going to
analyze in some detail, in the present paper,
the vertex function and its momentum-frequency
dependence in the superconducting 
state. Our conclusions show that, as soon as a superconducting
gap is opened, the momentum and frequency structure of the vertex
corrections is drastically changed, driving the system towards 
an effective strengthening of the electron-phonon coupling.

The paper is organized as follows:
in the next section we introduce the vertex function in the normal state
while in Sec.\ref{super} we study the modifications induced by the opening
of a superconducting gap. Section \ref{disc} is devoted to a general
discussion on the possible implications on physical quantities.

\section{The vertex function in the normal state}
\label{normal} 

The lowest order correction due to the breakdown of Migdal's 
theorem in the normal state is diagrammatically depicted in
Fig. 1, which defines the vertex function
$\Gamma({\bf q},\omega_n;{\bf k},\omega_{m})$:
\beginwide
\begin{equation}
\Gamma({\bf q},\omega_n;{\bf k},\omega_{m}) =\sum_{{\bf p}} 
\frac{2 |g({\bf k-p})|^{2}}{\omega({\bf k-p})} T \sum_{l}
\frac{\omega({\bf k-p})^{2}}
{(\omega_{m}-\omega_{l})^{2}+\omega({\bf k-p})^{2}}
\frac{1}{i\omega_{l}-\epsilon({\bf p})}
\frac{1}{i\omega_{l}+i\omega_n-
\epsilon({\bf p + q})} \:.
\label{nvertex}
\end{equation}
\endwidebottom
Here, $\epsilon({\bf p})$ and $\omega({\bf k}-{\bf p})$ are
electronic and phononic dispersions and $g({\bf k-p})$ is the
matrix element of the electron-phonon interaction, $T$ is the temperature
and $\omega_n$, $\omega_m$, and $\omega_l$ are
Matsubara frequencies.

As it was discussed in Ref.\onlinecite{PSG},
the qualitative behavior
of the momentum-frequency structure of 
$\Gamma({\bf q},\omega_n;{\bf k},\omega_{m})$,
is well caught by two characteristic
limits, the static and the dynamical one, $\Gamma_{s}$ and $\Gamma_{d}$,
defined as follows: 
$\Gamma_{s}=\Gamma({\bf q}\rightarrow 0,\omega_n=0;{\bf k},\omega_{m})$,
$\Gamma_{d}=\Gamma({\bf q}= 0,\omega_n\rightarrow 0;{\bf k},\omega_{m})$.
In fact, it can be show that the difference between
these two limits is in general finite and it is given by the
following expression:\cite{nota1}
\beginwide
\begin{equation}
\label{diff1}
\Gamma_d-\Gamma_s = \sum_{{\bf p}}
\frac{2 |g({\bf k-p})|^{2}}{\omega({\bf k-p})}
\frac{\partial f[\epsilon({\bf p})]}{\partial \epsilon({\bf p})} 
\frac{\omega({\bf k-p})^2}
{[\epsilon({\bf p})+\omega({\bf k-p})-i\omega_m]
[\epsilon({\bf p})-\omega({\bf k-p})-i\omega_m]}.
\end{equation}
\endwide
An analytic calculation of the vertex function $\Gamma$,
valid for small ${\bf q}$,
was also derived by considering an Holstein model, with
Einstein phonon dispersion $\omega({\bf q}) = \omega_{0}$,
a constant electron-phonon bare vertex 
$g({\bf q}) = g$,
and a flat electron density of states for a band of width $E$.
Within this model, at zero temperature and for a negligible external
frequency $\omega_m$, Equation (\ref{diff1}) simplifies to:

\begin{equation}
\label{diff2}
\Gamma_d-\Gamma_s=\lambda, 
\end{equation}
where $\lambda=2 g^2 N(0)/\omega_0$ is the electron-phonon coupling
constant and $N(0)$ is the electronic density of states at the Fermi level.
The two limits read 

\begin{equation}
\label{sdnorm}
\begin{array}{l}
\Gamma_{d}=\lambda/(1+2\omega_0/E) \\
\Gamma_{s}=-\lambda/(1+E/2\omega_0)\:,
\end{array}
\end{equation}
where $2\omega_0/E$ represents, in this model, the adiabatic parameter of Migdal's theorem.

The sign of the vertex function is plotted in Fig. 2
as function of the exchanged momentum $q$ and frequency 
$\omega_n$.
The non-analyticity of the vertex
at $({\bf q} = 0, \omega_n = 0)$
is here clear, where the negative static limit is recovered
on the vertical axis, and the positive dynamical limit on the
horizontal one. This complex structure in momentum and frequency
has important consequences on the role of the vertex 
corrections for different quantities. For instance, the
renormalized electron mass $m^{*}$, and so its isotope effect,
is mainly related to the static limit only\cite{GCP}, 
while for the
superconducting critical temperature the full
$({\bf q}, \omega_n)$ dependence of the vertex function is
relevant.\cite{GPSprl} 
This difference can be reflected in a stronger
dependence of $T_{c}$, and its isotope effect, on the microscopical parameters than the isotope effect on $m^{*}$,
as it is experimentally seen.\cite{zhao,muller}

However, we would like to stress that the above 
predictions are derived in the normal state.
How are these results affected by the onset of the 
superconductivity? 
Despite the difficulty to build a generalization of 
Migdal-Eliashberg theory including the nonadiabatic 
corrections to the Migdal's theorem {\em in the superconducting
state}, some preliminary answers at this question can be obtained
by analyzing the modifications on the vertex function 
induced by the superconducting order parameter.

\section{The vertex function in the superconducting state}
\label{super}

The existence of a long-range superconducting order parameter
modifies the vertex correction in two ways (Fig. 3). 
On one hand, the single particle propagators involved in 
the diagram (Fig. 3a) are affected by the opening of the 
superconducting gap. On the other hand, a new diagram appears,
involving two anomalous Green's functions (Fig. 3b). We 
label them $\Gamma^{on}$ and $\Gamma^{off}$, respectively, 
reminding that they are related to on-diagonal and off-diagonal
propagators.
In order to preserve the analogy
with the previous calculations at $T > T_{c}$ (Fig. 1), 
$\Gamma^{on}$ and $\Gamma^{off}$ are obtained
by using Green's functions renormalized only by
anomalous self-energies, and the normal state renormalization
is not taken in account. Within this approximation, the normal
and anomalous propagators are given by:

\begin{equation}
G({\bf k},i\omega_n)= 
-\frac{i\omega_n+\epsilon({\bf k})}
{\omega_n^{2}+E({\bf k})^2}
\label{Gfun}
\end{equation}
\begin{equation}
\:\: F({\bf k},i\omega_n)= 
\frac{\Delta}
{\omega_n^{2}+E({\bf k})^{2}} \:,
\label{Ffun}
\end{equation}
where $E({\bf k}) = \sqrt{\epsilon({\bf k})^{2}+\Delta^{2}}$
is the BCS superconducting excitation spectrum and $\Delta$ is
the order parameter. In this paper we consider only isotropic $s$-wave
symmetry, however most of the results will be valid also for
anisotropic $s$-wave and $d$-wave type of symmetries. 
By using the propagators defined above, the extension of 
Eq.(\ref{nvertex}) in the superconducting state reads:
\beginwide
\begin{eqnarray}
\Gamma({\bf q},\omega_n;{\bf k},\omega_{m}) & = & 
\Gamma^{on}({\bf q},\omega_n;{\bf k},\omega_{m}) +
\Gamma^{off}({\bf q},\omega_n;{\bf k},\omega_{m}) = 
\nonumber\\
\mbox{} & = & 
\sum_{{\bf p}} 
\frac{2 |g({\bf k-p})|^{2}}{\omega({\bf k-p})} T \sum_{l}
\left[ \frac{\omega({\bf k-p})^{2}}
{(\omega_{m}-\omega_{l})^{2}+\omega({\bf k-p})^{2}}\right] 
\frac{i\omega_{l}+\epsilon({\bf p})}
{\omega_{l}^{2}+E({\bf p})^2}
\frac{i\omega_{l}+i\omega_n+\epsilon({\bf p + q})}
{(\omega_{l}+\omega_n)^{2}+
E({\bf p + q})^2} +\nonumber\\
\mbox{} & \mbox{} & + 
\sum_{{\bf p}} 
\frac{2 |g({\bf k-p})|^{2}}{\omega({\bf k-p})} T \sum_{l}
\left[ \frac{\omega({\bf k-p})^{2}}
{(\omega_{m}-\omega_{l})^{2}+\omega({\bf k-p})^{2}}\right] 
\frac{\Delta}
{\omega_{l}^{2}+E({\bf p})^2}
\frac{\Delta}
{(\omega_{l}+\omega_n)^{2}+
E({\bf p + q})^2}. 
\label{svertex}
\end{eqnarray}
\endwidebottom

The first important difference of the above expression with respect to the 
vertex function in the normal state, Eq.(\ref{nvertex}), is already
evident by looking at the difference between the dynamical and static limits:

\beginwide
\begin{equation}
\label{diff3}
\Gamma_d-\Gamma_s=\sum_{{\bf p}}
\frac{2 |g({\bf k-p})|^{2}}{\omega({\bf k-p})}
\frac{\partial f[E({\bf p})]}{\partial E({\bf p})}
\frac{\omega({\bf k-p})^2\left[E({\bf p})^2-\omega({\bf k-p})^2
-\omega_m^2+2i\omega_m \epsilon({\bf p})\right]}
{[\omega({\bf k-p})^2+\omega_m^2-
E({\bf p})^2]^2+[2\omega_m E({\bf p})]^2}.
\end{equation}
\endwide
In fact, contrary to Eq.(\ref{nvertex}), the right hand side of 
Eq.(\ref{diff3}) vanishes when $T\rightarrow 0$ 
because of the gap opening in the quasiparticle spectrum. 
One can realize such a result by noticing that 
for $T\ll\Delta$ the integral over the Brillouin zone has roughly the
weight factor

\begin{equation}
\label{weight}
\frac{\partial f[E({\bf p})]}{\partial E({\bf p})}
\sim -\frac{\exp(-\Delta/T)}{T},
\end{equation}
which goes to zero for $T\rightarrow 0$. The vanishing of
$\Gamma_d-\Gamma_s$ at zero temperature 
is an intrinsic feature related to the opening of a gap in the
excitation spectrum. It is easy to verify that such a result
holds true also for a $d$-wave symmetry of the order parameter.
Therefore the equality $\Gamma_d=\Gamma_s$ at zero 
temperature characterizes the deep modification of the 
vertex correction due to the onset 
of superconducting long-range order.

In order to compare in more details 
the effect of the superconducting gap opening
on the vertex function 
we consider the same model discussed in
Ref.\onlinecite{PSG},
that we remind here briefly: the phonon dispersion is taken as a
Einstein one $\omega({\bf q}) = \omega_{0}$, the electron-phonon 
matrix element constant
$g({\bf q}) = g$, the electronic density of states flat:
\[
N(\epsilon) = N(0) \hspace{46pt} -\frac{E}{2} < \epsilon
< \frac{E}{2}\:,
\]
where $E$ is the total bandwidth and $E/2 =E_{F}$ 
the Fermi energy.
Moreover, since we are considering the 
vertex corrections at the first order in the adiabatic
parameter $\omega_{0}/E_{F}$, the sum on the internal
momentum ${\bf p}$ can be replaced by:\cite{PSG}
\[
\sum_{{\bf p}} \rightarrow N(0) \int^{E/2}_{-E/2} d\epsilon
\ll \ldots \gg_{FS} \:.
\]
For the same reason, $\epsilon({\bf p + q})$ can be written as
$\epsilon({\bf p + q}) \simeq \epsilon({\bf p}) +
v_{F} |{\bf q}| \cos\theta = \epsilon + y$, and the variable
$y$ represents the angular degrees of freedom to be averaged
on the Fermi surface. In this context, the vertex function will
depend on momenta just via
the dimensionless variable $Q= |{\bf q}|/2 k_{F}$, where
$|{\bf q}|$ is the momentum transfer. Finally, since
the vertex function 
$\Gamma({\bf q},\omega_n;{\bf k},\omega_{m})$ 
depends weakly on the external frequency 
$\omega_{m}$, we set $\omega_{m} = 0$. 
With these notations, the expression of the vertex function 
$\Gamma$ is:
\beginwide
\begin{equation}
\Gamma(Q,\omega_n) = \frac{\lambda}{EQ}
T \sum_{l}
\left[
\frac{\omega_{0}^{2}}
{\omega_{l}^{2}+\omega_{0}^{2}} \right]
\int_{-E/2}^{E/2} d\epsilon \int_{-EQ}^{EQ} dy 
\frac{\epsilon(\epsilon + y) -
\omega_{l}(\omega_n+\omega_{l}) +\Delta^{2}}
{\left[ \omega_{l}^{2}+\epsilon^{2}+\Delta^{2} \right]
\left[(\omega_{l}+\omega_n)^{2}+
(\epsilon + y)^{2}+\Delta^{2}\right]}\:,
\label{svertexsimple}
\end{equation}
\endwide
where, as before, $\lambda= 2g^{2}N(0)/\omega_{0}$.
The sum over $\omega_{l}$ can be performed exactly
by using the Poisson's formula at $T = 0$ : 
$T \sum_{l} \rightarrow 
\int_{-\infty}^{\infty} d\omega /2\pi$. 
In this way, the static and dynamical limits can be analytically
obtained. The derivation is quite cumbersome, but 
straightforward, and does not present particular difficulties.
Namely, we obtain
\beginwide
\begin{eqnarray}
\label{sdsup}
\Gamma_{s}=
\Gamma_{d} & = & \lambda \int_{-E/2}^{E/2} 
\frac{\omega_{0} d\epsilon}{2(\omega_{0}+
\sqrt{\epsilon^{2}+\Delta^{2}})^{2}} \nonumber \\
& = & \lambda
\left[\frac{\delta^2}{4[1-\delta^2]^{3/2}}
\ln\left|
\frac{[(\sqrt{1+\delta^2m^2}-\sqrt{1-\delta^2})^2-\delta^4m^2]}
{[(\sqrt{1+\delta^2m^2}+\sqrt{1-\delta^2})^2-\delta^4m^2]}
\frac{[m\sqrt{1-\delta^2}+1]^2}{[m\sqrt{1-\delta^2}-1]^2}
\right|+\frac{m-\sqrt{1+\delta^2m^2}}
{[1-\delta^2][m^2(1-\delta^2)-1]}\right],
\end{eqnarray}
\endwide
where, for the sake of shortness, we have set 
$m =2\omega_0/E$ and  $\delta = \Delta/\omega_0$.
Comparing Eq.(\ref{sdsup}) with Eq.(\ref{sdnorm}) we
can see that the onset of the long-range
superconducting order parameter leads to a drastic
change of the structure of the vertex function. In particular,
while the dynamical limit is affected in a smooth way by
a finite $\Delta$, the value of the static limit
immediately jumps to the positive value
$\Gamma_{s} = \Gamma_{d}$ as soon as the gap is open.
This result is not surprising, since similar modifications
of the analytical properties of different susceptibilities
(and the electron-phonon vertex can be described as a particular
susceptibility) were already recovered. \cite{nam,scalapino} 

The global change of $\Gamma(Q,\omega_n)$ due to the opening
of the superconducting gap is also more evident in Fig. 4, where 
we plot the sign of the vertex function 
in the $Q-\omega_n$ space, as calculated directly by 
Eq. (\ref{svertexsimple}), without any small $Q$ assumption.
With respect to the normal state, 
marked by the dashed line in Fig. 4,
the change of sign in the superconducting state
is shifted towards larger values of $Q$,
leading to an enlargement of the region of positivity of the vertex function.
Moreover, the  non-analyticity
of the vertex function in the normal state at the point
$({\bf q}=0,\omega_n=0)$ is completely removed.
For zero exchanged frequencies $\omega_n=0$, 
the vertex changes sign at a finite value $Q_\Delta$
of the exchanged momentum,
roughly given by $Q_\Delta \sim \Delta / \omega_0$.
This feature can be attributed to the presence of a characteristic 
length which we identify with the superconducting coherence 
length $\xi_0$. The interplay of the vertex
function and the coherence length will be discussed from
the physical point of view in Sec.\ref{disc}.

In addition to the increase of the region where $\Gamma(Q,\omega_n)>0$,
the opening of the superconducting gap leads also to an amplification
of the global magnitude of the vertex. This behavior is clearly shown in
Fig. 5, where we plot the average of the vertex function over the Brillouin zone
and over the relevant exchanged frequencies:\cite{PSG,CP}
\begin{equation}
\label{average}
\langle \Gamma \rangle =
\int_{-\infty}^{\infty} \frac{d\omega}{\pi} 
\frac{\omega_0}{\omega^2+\omega_0^2} 
\int_0^1\! 2Q \Gamma(Q,\omega)dQ .
\end{equation}
When $\Delta=0$ the positive and negative contributions of the
vertex function nearly cancel each other giving a negligible
total effect also for intermediate values of $\omega_0/E_F$. 
The case $\Delta\neq 0$ gives instead an
imbalance in favour of the positive contribution resulting
into an enhancement of $\langle \Gamma \rangle >0$. This result
is a consequence of both the enlargement of the positive
sign region and at the same time an increase (suppression)
of the magnitude of the positive (negative) values of $\Gamma(Q,\omega_n)$.
Despite of the deep modifications induced by the opening of
gap, the vertex function still fulfills the adiabatic limit
of Migdal's theorem 
$\lim_{\omega_0/E_F\rightarrow 0} \langle \Gamma \rangle=0$. 

The importance of small momentum transfer selection for
the global sign of the vertex function in the normal state
has already be outlined in Refs. \onlinecite{PSG,GPSprl}.
There, an upper cut-off $Q_c$ was introduced in order
to select values of the momentum transfer $Q<Q_c$. In this way
the average procedure of Eq.(\ref{average}) was modified to restrict
the integration over $Q$ according to the constraint $Q<Q_c$.
The so defined constrained average turns out to
be very sensitive to $Q_c$ since it modifies the balance of
the positive and negative regions of $\Gamma(Q,\omega_n)$. In
particular, small values of $Q_c$ lead to an effective enhancement
of $\langle \Gamma \rangle$.
This situation is quite altered by the opening of a finite gap.
In fact, from Fig. 4 it can be realized that an upper cut-off $Q_c$
has less important effects on the global magnitude when $\Delta\neq 0$.
Therefore, in the superconducting state,  the $Q_c$ dependence
of $\langle \Gamma \rangle$ is more and more weakened by increasing 
the value of $\Delta$. This feature is described in Fig. 6 where
we compare $\langle \Gamma \rangle$ for $\Delta=0$ and 
$\Delta=0.5\omega_0$ as a function of $Q_c$.
However, more than the magnitude of $\langle \Gamma \rangle$,
the most significant quantity to look at
is the {\em sensitivity} of the average vertex, 
that can be expressed as the derivative
$\partial \langle \Gamma \rangle / \partial Q_c$ 
(insert of Fig. 6).
For small $Q_c$'s, that in our perspective is the relevant region
for the high-$T_c$ materials, the sensitivity of the
averaged vertex function $\langle \Gamma \rangle$ is much less
pronounced for $\Delta\neq 0$, implying that 
microscopical variations of $Q_c$ lead to smaller
modifications of $\langle \Gamma \rangle$ in the 
superconducting state than in the normal one.

As we are going to discuss below, all these modifications on the
vertex structure induced by the gap opening can have important
implications on physical quantities of the superconducting state.

\section{Discussion}
\label{disc}

As we have seen in the above calculations, the onset
of a long-range superconducting order parameter
changes substantially the momentum-frequency structure 
of the vertex function. In particular, the region of positivity
is enlarged with respect the normal state, 
and its relevance for small adiabatic parameter is increased.
The main modifications can be sketched by the static and 
dynamical limits. The dynamical limit $\Gamma_d$ is just slightly
affected by the onset of superconductivity, and it remains 
always positive, whereas the static limit $\Gamma_s$ is drastically
changed switching from a negative to a positive value. 
In particular, at zero temperature these two limits are strictly
equal for $\Delta\neq 0$.
This behavior can be understood by 
considering that the dynamical limit, which is essentially
independent of the many-particle properties, is
related to the static lattice distortions, and as matter of 
fact it defines the attractive electron-phonon coupling. 
On the other hand,
the static limit depends on many-body effects due to the
Pauli exclusion principle, so that 
it is strongly determined by the 
fermionic picture of the single-particle properties of the
electrons.\cite{GPM} Such a  picture is completely
destroyed in the superconducting state where all the Fermi 
electrons are condensed in Cooper pairs,
and Pauli principle is ineffective. As a result, no more 
difference between dynamical and static limits is found.
Moreover, the presence of a finite value $Q_{\Delta}$
of the momentum transfer $Q$ for which at zero exchanged
frequency the vertex changes sign is a consequence of
the presence of the superconducting coherence length $\xi_0$.

Therefore, based on this result, which kind of physical 
predictions can be done?

First of all, it is straightforward to expect that the
vertex corrections beyond Migdal's theorem become more and
more important in the superconducting state with the
magnitude of the gap, and, since the positive part
of the vertex function is increased, this leads to
a strengthening of the electron-phonon coupling.
We remind that the magnitude of the superconducting gap
$\Delta(0)$ is of the same order of the phonon frequencies in
the high-$T_{c}$ materials\cite{plakida}, 
so that such a situation
gives a huge increase of the positive part of the 
vertex function.
As a first consequence, the zero temperature quantities,
as for instance $\Delta(0)$, will feel a much stronger effect
of vertex corrections with respect $T_{c}$, and the ratio
$2 \Delta / T_{c}$ is expected to increase the 
BCS value $3.53$. The modifications induced by 
the superconducting gap on the vertex corrections are 
essentially ruled by  ratio $\Delta(0)/\omega_0$, 
so that materials with high
phonon energies, as the $C_{60}$ compounds\cite{gunnarson},
should show a lighter enhancement of $2 \Delta / T_{c}$ 
with respect the cuprates, where the phonon energies 
involved are smaller.\cite{plakida}
This is actually recovered by experimental 
measurements.\cite{koller,batlogg}

In addition we can argue that the superconducting properties
are much more robust {\em inside} the superconducting phase
than the properties related to the onset of superconducting instabilities. 
In fact, as the positive and negative parts
of the vertex function are almost equivalent in the normal state,
the critical temperature is strongly affected by the an 
implicit possible predominance of forward 
small ${\bf q}$ scattering, that
will select the positive part. Such
a situation is naturally given by taking in account the
strong electronic correlation due to the on-site Coulomb
interaction\cite{zeyher}, that
is considered to be a crucial ingredient of the high-$T_{c}$ superconductors. 
The relevance of the forward scattering is strongly affected
by microscopic parameters, like
the adiabatic ratio and
the doping. As a consequence, the critical temperature itself is expected to be  strongly related to these parameters.
On the other hand, 
due to the opening of the superconducting gap, the vertex 
corrections involved in the zero temperature gap $\Delta(0)$
are essentially positive, and the small ${\bf q}$ selection
given by the strong correlation is less relevant.
This could explain the different behavior of 
$\Delta(0)$ and $T_{c}$ as function of the doping $\delta$.
\cite{loram}

\acknowledgments
We thank S. Ciuchi and M. Scattoni for useful discussions.
C. G. acknowledges the support of a I. N. F. M. PRA project.

{\Large {\bf Figures:}}

\begin{figure}
\centerline{\psfig{figure=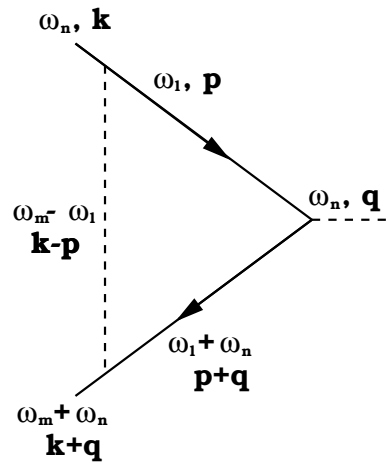,width=5cm}}
\narrowtext
\caption{The lowest order vertex correction diagram. Solid lines
are electronic propagators, dashed lines are phonon 
propagators.}
\label{vertnor2}
\end{figure}

\begin{figure}
\centerline{\psfig{figure=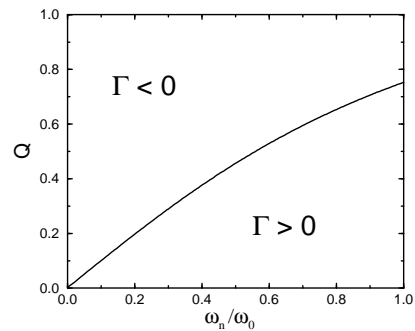,width=7cm}}
\narrowtext
\caption{Positive and negative regions of the vertex function
in the normal state (Ref. 4). It was derived by a small Q expansion. The adiabatic parameter is 
$2\omega_{0}/E = 0.5$.}
\label{cut0}
\end{figure}

\begin{figure}
\centerline{\psfig{figure=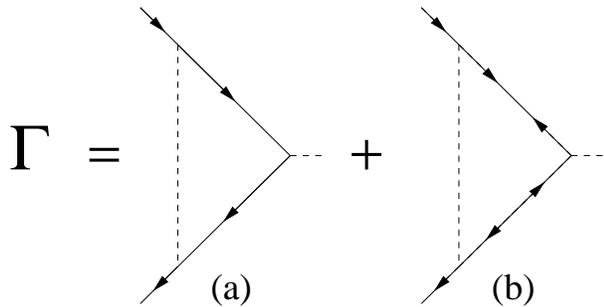,width=8cm}}
\narrowtext
\caption{First order vertex diagram in the superconducting state.}
\label{vertsup}
\end{figure}

\begin{figure}
\centerline{\psfig{figure=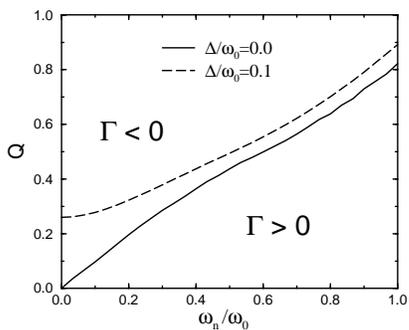,width=7cm}}
\narrowtext
\caption{Comparison between positive and negative 
regions of the vertex function
in the normal state ($\Delta =0$) and in the superconducting state ($\Delta =0.1\omega_0$) for $2\omega_{0}/E = 0.5$.}
\label{cut01}
\end{figure}

\begin{figure}
\centerline{\psfig{figure=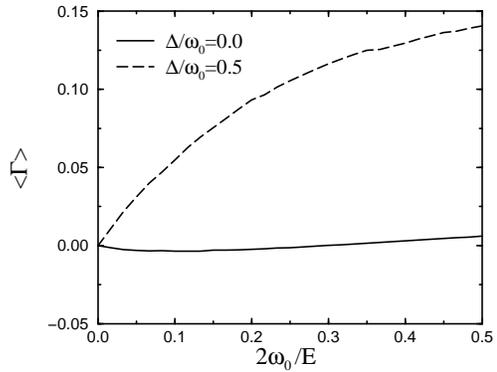,width=8cm}}
\narrowtext
\caption{Momentum-frequency average of the vertex function
$\langle \Gamma(Q,\omega) \rangle$ as function of the adiabatic
parameter $2\omega_{0}/E$ for different gap magnitudes.
$\lambda$ is set equal 1.}
\label{media}
\end{figure}

\begin{figure}
\centerline{\psfig{figure=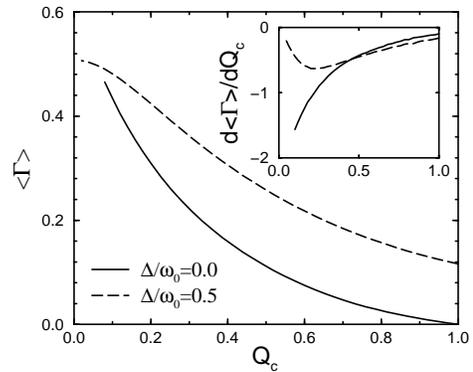,width=8cm}}
\narrowtext
\caption{Plot of $\langle \Gamma \rangle$ 
as function of the momentum cut-off $Q_c$,
for $2\omega_{0}/E = 0.3$ and $\lambda = 1$. 
Insert: the corresponding derivative:
$\partial \langle \Gamma \rangle / \partial Q_c$.}
\label{qc}
\end{figure}

\end{multicols}

\begin{references}


\bibitem{plakida}
N.M. Plakida, {\em High temperature superconductivity:
experiment and theory}, Springer, Berlin (1995).

\bibitem{migdal}
 A.B. Migdal, Sov. Phys. JETP {\bf 34}, 996 (1958).

\bibitem{schrieffer}
J.R. Schrieffer, J. Low Temp. Phys. {\bf 99}, 377 (1995).

\bibitem{PSG}
L. Pietronero, S. Str\"{a}ssler and C. Grimaldi, 
Phys. Rev. B {\bf 52}, 10516 (1995);
C. Grimaldi, L. Pietronero and S. Str\"{a}ssler, 
Phys. Rev. B {\bf 52} 10530 (1995).

\bibitem{GPSprl} 
C. Grimaldi, L. Pietronero and S. Str\"{a}ssler, 
Phys. Rev. Lett. 
{\bf 75}, 1158 (1995).

\bibitem{CP} 
E. Cappelluti and L. Pietronero, Phys. Rev. B 
{\bf 53}, 932 (1996).

\bibitem{GCP} 
C. Grimaldi, E. Cappelluti and L. Pietronero, 
preprint cond-mat/9710159.

\bibitem{zhao} G.M. Zhao and D.E. Morris, Phys. Rev. B 
{\bf 51}, 16487 (1995).

\bibitem{muller} G.M. Zhao, M.B. Hunt, H. Keller 
and K.A. M\"uller, Nature 
{\bf 385}, 236 (1997).

\bibitem{freericks} 
Lowest order corrections beyond Migdal's theorem
to Eliashberg equations  for $T < T_c$ were analyzed in
J. K. Freericks, E. J. Nicol, A. Y. Liu,
and A. A. Quong, Phys. Rev. B {\bf 55}, 11651 (1997).
However, a local assumption
was there used, so that the crucial dependence on the 
exchanged momenta was neglected.

\bibitem{nota1}
Strictly speaking, this is true only for a finite electron
density. In the one-electron case, the difference
between the dynamical and static limits becomes zero.


\bibitem{nam} S.B. Nam, Phys. Rev.  
{\bf 156}, 470 (1967).

\bibitem{scalapino} D.J. Scalapino, S.R. White and
S.C. Zhang, Phys. Rev. B 
{\bf 47}, 7995 (1993).

\bibitem{GPM} C. Grimaldi, L. Pietronero and M. Scattoni,
unpublished.

\bibitem{gunnarson} O. Gunnarsson, Rev. Mod. Phys. {\bf 69},
575 (1997).

\bibitem{koller} D. Koller {\em et al.},
Phys. Rev. Lett.
{\bf 77}, 4082 (1996).

\bibitem{batlogg} B. Batlogg,
Physica B
{\bf 169}, 7 (1991).

\bibitem{zeyher} R. Zeyher and M. Kuli\'c
Phys. Rev. B 
{\bf 53}, 2850 (1996).

\bibitem{loram} J.W. Loram, K.A. Mirza, J.M. Wade,
J.R. Cooper and W.Y. Liang,
Physica C {\bf 235-240}, 134 (1994).

\end{references}
\end{document}